\documentclass[english,twocolumn,prl,amssymb,amsmath,superscriptaddress,showpacs]{revtex4}
\usepackage[latin9]{inputenc}
\usepackage{graphicx,color}
\usepackage[bookmarks]{hyperref}
\usepackage{babel}

\newcommand{\be}{\begin{equation}}
\newcommand{\ee}{\end{equation}}
\newcommand{\ua}{\uparrow}
\newcommand{\da}{\downarrow}
\newcommand{\ch}{\mathrm{ch}}
\newcommand{\sh}{\mathrm{sh}}
\newcommand{\tah}{\mathrm{th}}

\begin{document}

\title{Scaling and Universality at Dynamical Quantum Phase Transitions}

\author{Markus Heyl}
\affiliation{Institute for Quantum Optics and Quantum Information of the Austrian Academy of Sciences, 6020 Innsbruck, Austria}
\affiliation{Institute for Theoretical Physics, University of Innsbruck, 6020 Innsbruck, Austria}
\affiliation{Physik Department, Technische Universit\"at M\"unchen, 85747 
Garching, Germany}

\begin{abstract}

Dynamical quantum phase transitions (DQPTs) at critical times appear as non-analyticities during nonequilibrium quantum real-time evolution. Although there is evidence for a close relationship between DQPTs and equilibrium phase transitions, a major challenge is still to connect to fundamental concepts such as scaling and universality.
In this work, renormalization group transformations in complex parameter space are formulated for quantum quenches in Ising models showing that the DQPTs are critical points associated with unstable fixed points of equilibrium Ising models. Therefore, these DQPTs obey scaling and universality. On the basis of numerical simulations, signatures of these DQPTs in the dynamical buildup of spin correlations are found with an associated power-law scaling determined solely by the fixed point's universality class. An outlook is given on how to explore this dynamical scaling experimentally in systems of trapped ions.

\end{abstract}
\date{\today}
\maketitle

\emph{Introduction:--} In equilibrium phase transitions are of fundamental importance both for the theoretical understanding of physical systems as well as for applications. In this context, continuous phase transitions are of particular interest because they exhibit scaling and universality~\cite{Sachdev2011}. These fundamental concepts are intimately connected to renormalization group (RG) theory and the associated fixed points. 
Dynamical quantum phase transitions (DQPTs) during quantum real-time evolution have emerged as a nonequilibrium analogue to equilibrium phase transitions where Loschmidt amplitudes
\be
	\mathcal{G}(t) = \langle \psi_0 | e^{-i H t} | \psi_0\rangle
\label{eq:defLoschmidtAmplitude}
\ee
become nonanalytic as a function of time~\cite{Heyl2013a}. Here, $|\psi_0\rangle$ is an initial pure state and $H$ the Hamiltonian driving the coherent time evolution. By now, DQPTs have been discovered in a variety of different contexts~\cite{Heyl2013a,Heyl2014,Fagotti2013,Karrasch2013,Kriel2014,Andraschko2014,Hickey2014,Canovi2014fo,Vajna2014,Vajna2014tc,James2015,Budich2015} and indications for a close relationship between DQPTs and equilibrium phase transitions have been found~\cite{Heyl2013a,Andraschko2014,Heyl2014,Canovi2014fo,Kriel2014,Karrasch2013}. But still, a major challenge is to connect to fundamental concepts such as scaling and universality.

In this work it is shown for the first time that DQPTs obey scaling and universality. For that purpose, Loschmidt amplitudes for quantum quenches in one- and two-dimensional Ising models are mapped exactly onto equilibrium partition functions at complex couplings for which RG transformations in complex parameter space are formulated. As a main result, DQPTs are critical points on the attracting manifold of the unstable fixed points of this RG. Therefore, Loschmidt amplitudes satisfy a scaling form with exponents determined solely by the underlying universality class. Moreover, numerical evidence is provided that the critical phenomena in Loschmidt amplitudes are related to dynamical power-law scaling in spin-spin correlations. An outlook is given on how to verify this scaling experimentally in systems of trapped ions within current technology.

\begin{figure}
\centering
\includegraphics[width=\linewidth]{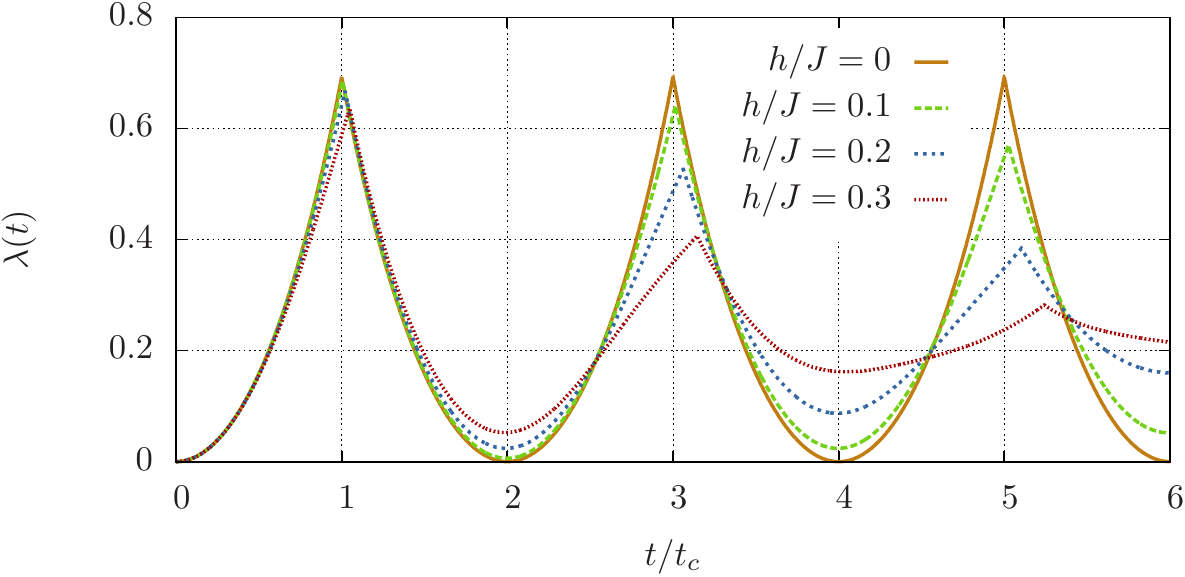}
\caption{(color online) Dynamical quantum phase transitions in the Loschmidt echo rate function $\lambda(t) = -N^{-1} \log [|\mathcal{G}(t)|^2]$ after quenches in the 1D Ising chain. The nonanalytic kink structure of $\lambda(t)$ is a direct consequence of the universal exponents of the underlying fixed point, see Eq.~(\ref{eq:gScaling1D}).}
\label{fig:1}
\end{figure}

Universality and scaling of DQPTs will be studied for quantum quenches in transverse-field Ising models:
\be
	H(h) = -\sum_{\langle lm \rangle} J_{lm} \sigma_l^z \sigma_m^z - h\sum_{l=1}^L \sigma_l^x.
\label{eq:IsingHamiltonian}
\ee
with $\sigma_l^\alpha$, $\alpha=x,z$, Pauli matrices on lattice site $l = 1,\dots,L$ and $L$ the total number of spins. While in one dimension (1D) the nearest-neighbor (NN) coupling $J_{lm}>0$ is taken as uniform $J_{lm}=J$, in two dimensions (2D) an anisotropic square lattice is considered with couplings $J$ within the rows and $J_\perp$ along the columns.

The Ising model supports DQPTs both in 1D~\cite{Pollmann2010dv,Heyl2013a,Vajna2014} and also in 2D~\cite{James2015}. Fig.~\ref{fig:1} shows DQPTs for quantum quenches in the 1D case. Throughout this letter, plots of the DQPTs will be given in terms of the Loschmidt echo $\mathcal{L}(t) = |\mathcal{G}(t)|^2$ quantifying the magnitude of $\mathcal{G}(t)$. Specifically, due to the large-deviation scaling of $\mathcal{L}(t)$~\cite{Silva2008gj,Gambassi2012a,Heyl2013a} it is suitable to introduce its rate function
\be
	\lambda(t) = -\frac{1}{N} \log[\mathcal{L}(t)],
\ee
which, in contrast to $\mathcal{L}(t)$, is intensive in the thermodynamic limit. As in Fig.~\ref{fig:1}, quantum quenches from fully polarized initial states
\be
|\psi_0\rangle = \bigotimes_{l=1}^L |\rightarrow \rangle_l\,\, |\rightarrow\rangle_l = \frac{1}{\sqrt{2}} \left[ |\ua\rangle_l + |\da \rangle_l \right]
\label{eq:initialState}
\ee
will be considered, i.e., ground states of initial Hamiltonians $H_0 = H(h\to\infty)$.

In the following, it is the aim of the present work to relate the DQPTs and therefore the nonanalytic structure of $\lambda(t)$ to scaling in the vicinity of unstable fixed points. For that purpose, it will be shown that Loschmidt amplitudes in Eq.~(\ref{eq:defLoschmidtAmplitude}) can be mapped onto equilibrium partition functions of classical Ising models at complex couplings for the considered parameter regime. In order to address scaling and universality, real-space decimation RGs, exact in 1D and approximate in 2D, are formulated. The most important result of the analysis of the RG equations is that DQPTs are critical points flowing to unstable fixed points of equilibrium Ising models implying universality and scaling. Notice that singular behavior can also occur in the Fourier transform of the Loschmidt amplitude~\cite{Silva2008gj,Gambassi2012a,Heyl2012ys,Palmai2014} which, however, is of different nature.


\emph{Equilibrium partition functions.--} Let us first consider vanishing final transverse fields with a Hamiltonian $H(h=0)=-J\sum_{\langle lm \rangle} \sigma_l^z \sigma_m^z$. Then, as will be outlined below, the Loschmidt amplitude $\mathcal{G}(t)$ becomes:
\be
	\mathcal{G}(t) = \frac{1}{2^L} \mathrm{Tr} \left[ e^{it\sum_{\langle l m \rangle} J_{lm} \sigma_l^z \sigma_m^z }\right],
\label{eq:loschmidtPartitionFunction}
\ee
with $\mathrm{Tr}$ denoting the trace over Hilbert space. Remarkably, this is nothing but the equilibrium partition function of a classical Ising model at complex coupling $iJ$, inverse temperature $t$, and $g(t)=-N^{-1}\log[\mathcal{G}(t)]$ the associated free energy density (apart from an overall temperature normalization). The above identification follows by using two properties. First, $|\psi_0\rangle$ in Eq.~(\ref{eq:initialState}) can be written as an equally weighted superposition of all spin configurations in the $\sigma_l^z$ basis. Secondly, $H(h=0)$ does not induce spin flips and therefore only the diagonal matrix elements contribute which is nothing but the trace. In analogy to the equilibrium case it will be suitable to introduce dimensionless couplings
\be
	K = iJt, \quad K_\perp = iJ_\perp t,
\label{eq:defK}
\ee
which in the present nonequilibrium context, however, are now complex.  Notice that the above identification of $\mathcal{G}(t)$ with a partition function is independent of dimension. In the following, the 1D and 2D cases will be considered because they allow for exact solutions.


\emph{One dimension:--} The 1D Ising model can be solved on the basis of the transfer matrix $T$:~\cite{Sachdev2011}
\be
	\mathcal{G}(t) = \mathrm{tr} \, T^L, \,\,\,\, T = \frac{1}{2} \left( \begin{array}{cc} e^{K} & e^{-K} \\ e^{-K} & e^{K} \end{array} \right),
\label{eq:1DTmatrix}
\ee
with $\mathrm{tr}$ denoting the trace over a basis of the $2\times 2$ matrix problem. The matrix $T$ has two eigenvalues $\nu_c = \ch(K)$ and $\nu_s = \sh(K)$ with $\sh(K)$ and $\ch(K)$ the hyperbolic sine and cosine, respectively. For $L\to\infty$, $\mathcal{G}(t)$ is dominated by the eigenvalue of largest magnitude, i.e., $\mathcal{G}(t) = \nu^L$ with $\nu = \nu_c$ if $|\nu_c|>|\nu_s|$ and $\nu = \nu_s$ otherwise.
Notice, that $\nu$ can switch between $\nu_s$ and $\nu_c$ yielding a nonanalytic structure in $\lambda(t) = -2 \mathrm{Re} \left[\log(\nu)\right]$. This switching of the dominant eigenvalue is the underlying origin of the DQPTs in 1D Ising models, as has also been seen in XXZ chains~\cite{Andraschko2014}. The critical times $t_n$ of the DQPTs are given by the condition $|\nu_c|=|\nu_s|$, i.e., $t_n=\pi(2n+1)/(4J)$ with $n\in\mathbb{Z}$, compare also Ref.~\cite{Heyl2013a}. In equilibrium, $|\nu_c| = |\nu_s|$ can only be satisfied in the limit of zero temperature $T=0$. When discussing the anticipated RG procedure below it will be shown that this correspondence is not accidental but rather has a very profound origin.

Having established the presence of DQPTs it is now the aim to address the question of scaling and universality. For that purpose an RG scheme in complex parameter space will now be introduced. RG transformations in complex parameter space have been previously studied in the context of the standard model~\cite{Denbleyker2010ty,Liu2011ga} as well as for equilibrium  partition functions in complex parameter spaces~\cite{Damgaard1993,Wei2014pt}. Eliminating every second spin via decimation~\cite{Nelson1975zz,Damgaard1993,Wei2014pt} an exact RG transformation can be formulated yielding the following recursion relation~\cite{supp}
\be
	\tah(K') = \tah^2(K).
\label{eq:RGEq1D}
\ee
As a result, it is found that the RG has two fixed points $K^* = 0,\infty$ corresponding to the equilibrium ones at infinite and zero temperature  even when $K$ is complex initially. For $K$ with $|K|\ll 1$ this gives $K'=K^2$ implying that the fixed point $K^*=0$ is stable. For $K=K^* + \delta K$ in the vicinity of $K^*=\infty$ one obtains that $\delta K' = 2\, \delta K=b^\lambda \, \delta K$ with $b=2$ the change in length scale due to the decimation and $\lambda=1$ the associated anomalous dimension. Thus, it is found here that the fixed point $K^*=\infty$ is unstable as in the equilibrium case. But remarkably, this is not necessarily true for initial couplings beyond the linear regime. In particular, the DQPTs at times $t_n$ map onto the $K^*=\infty$ fixed point after precisely two RG steps. Times $t$ with weak deviation $\tau = (t-t_c)/t_c$ from a DQPT at $t_c$ map after two RG steps onto the linear regime of the unstable fixed point. Using that $\lambda=1$, one can then directly deduce the scaling form of $g(t)$:
\be
	g(\tau) \sim |\tau|^{d/\lambda} \Phi_{\pm} = |\tau| \Phi_{\pm}, \,\,\, \tau = \frac{t-t_c}{t_c}.
\label{eq:gScaling1D}
\ee
with dimension $d=1$ and $\Phi_\pm$ a constant that may differ for $\tau\lessgtr0$. Remarkably, this is indeed the scaling behavior that is found from the exact solution of $g(t)$ in the vicinity of the DQPTs~\cite{Heyl2013a}, compare also Fig.~\ref{fig:1}. Therefore, the nonanalytic behavior of $g(t)$ can now be attributed to an unstable fixed point of an RG allowing to extend fundamental concepts such as robustness, scaling, and universality to the nonequilibrium regime on general grounds. Notice that robustness has been established for particular cases recently~\cite{Karrasch2013,Kriel2014}. In the present work a specific initial state and final Hamiltonian has been considered so far. However, the identification of DQPTs with unstable fixed points allows to conclude that weak symmetry-preserving perturbations do not change the universal properties. In the following this will now be demonstrated by incorporating a weak transverse field in the final Hamiltonian.

\emph{Transverse Fields.--}  For $h/J \ll 1$ the field part $V=-h\sum_l \sigma_l^x$ of Eq.~(\ref{eq:IsingHamiltonian}) can be eliminated using standard time-dependent perturbation theory~\cite{supp}.
To first order in $h/J$ one obtains that $\mathcal{G}(t) =2^{-L} \mathrm{Tr} e^{\overline{H}}$ can be again represented in terms of an effective classical Ising model $\overline{H}$, but now including also next-to-nearest neighbor (NNN) interactions~\cite{supp}:
\be
	\overline{H} = K \sum_{l} \sigma_l^z \sigma_{l+1}^z + G \sum_{l} \sigma_l^z \sigma_{l+2}^z 
\ee
with $G=-iht/2 +i h\sin(4Jt)/(8J)$ and a modified coupling $K = iJt+h[1-\cos(4Jt)]/(4J)$. 
Within the same decimation RG as used for the NN case let us eliminate every second lattice site. Based on a cumulant expansion~\cite{Niemeyer1973tr} for the perturbative NNN couplings one obtains to first order in $G$:~\cite{supp}
\be
	K' = P+ G \left[1+\frac{1-e^{-4P}}{2} \right] ,\,\,\, G'  = G \,\, \frac{1-e^{-4P}}{4}
\ee
with $\tah(P) = \tah^2(K)$ the solution at $h=0$, see Eq.~(\ref{eq:RGEq1D}). This set of RG equations exhibits two fixed points $(K^*,G^*) = (0,0),(\infty,0)$. In the vicinity of the unstable fixed point $K^*=\infty$ we get $\delta K' = 2\delta K + 3 G/2$ and $G'=G/4$ and therefore weak fields $h/J \ll 1$ constitute an irrelevant perturbation. This is in perfect agreement with the exact solution where it can be seen that $h/J>1$ is necessary to destroy the DQPTs~\cite{Heyl2013a}, see also Fig.~\ref{fig:1}. Moreover, the scaling properties of the DQPTs are invariant under a slight modification of the initial state by taking the ground state for an initial transverse field $1<h/J <\infty$. Therefore, it is expected that the main results are also valid beyond the case of a fully polarized state studied here.


\begin{figure}
\centering
\includegraphics[width=\linewidth]{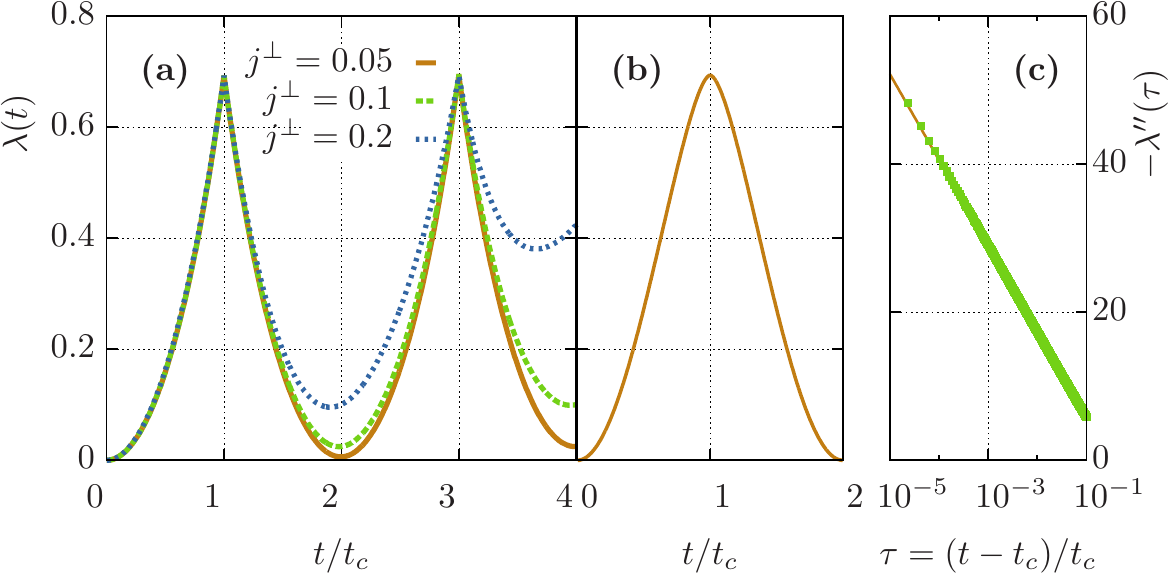}
\caption{(color online) Dynamics of $\lambda(t)$ in the 2D Ising model. (a) For strong anisotropies $j_\perp = J_\perp/J \ll 1$ the weak coupling $J_\perp$ represents an irrelevant perturbation and the critical properties, i.e., the kinks, are identical to the 1D case. (b) For the isotropic 2D Ising model the nonanalytic structure changes to a logarithmic singularity which is illustrated in the second derivative shown in (c).}
\label{fig:2}
\end{figure}
\emph{Two dimensions:--} The partition function of the 2D Ising model can be solved exactly~\cite{Onsager1944kb,Kaufman1949pp,Schultz1964qt}. For the case of complex $K$ this is still possible which yields:
\be
	g(t) = - \frac{1}{2}\log[2 \sinh(K)] -\int_{-\pi}^\pi \frac{dq}{4\pi} s(\varepsilon_q) \varepsilon_q
\label{eq:solutionPT2D}
\ee
with $\varepsilon_q$ the solution of the equation:
\be
	\mathrm{ch}(\varepsilon_q) = \mathrm{ch}(2K_\perp)\mathrm{ch}(2\overline{K}) - \mathrm{sh}(2K_\perp)\mathrm{sh}(2\overline{K}) \cos(q).
\ee
and $s(x) = \mathrm{sgn}[\mathcal{R}(x)]$ returns the sign of its argument's real part. Here, $\overline{K}$ is given by the condition $\tah(K) = \exp(-2\overline{K})$ with $\tah(K)=\sh(K)/\ch(K)$.

In Fig.~\ref{fig:2}, the dynamics of $\lambda(t)$ is shown for different anisotropies $j_\perp=J_\perp/J$. For $j_\perp \ll 1$, DQPTs are found at times $t_n=(2n+1)\pi/(4J)$ which are solely controlled by the coupling $J$. Indeed, it will be shown below using a perturbative RG that a weak coupling $j_\perp \ll 1$ represents an irrelevant perturbation. If, however, $j_\perp=1$ a drastic change in the nature of the DQPT occurs with a logarithmic nonanalyticity:
\be
	g(\tau) \sim \tau^2 \log(|\tau|),
\label{eq:gScaling2D}
\ee
which is illustrated in Fig.~\ref{fig:2}. Notice the remarkable similarity of the scaling behavior in Eq.~(\ref{eq:gScaling2D}) with the equilibrium free energy at the equilibrium critical point of the 2D Ising model when $\tau$ denotes the relative temperature distance to the critical point~\cite{Fisher1967da}. 

As opposed to 1D, it is not possible to derive a closed set of exact RG recursion relations for the 2D case. In the limit of strong anisotropy $j_\perp \ll 1$, however, an approximate RG can be constructed. For that purpose let us decompose the square lattice into even and odd rows. The odd rows can be eliminated perturbatively using a cumulant expansion~\cite{Niemeyer1973tr} which in the present case is controlled via $|K_\perp|\ll 1$. To second order in $K_\perp$ one obtains:~\cite{supp}
\begin{align}
	K' & = K + 2 Q K_\perp^2, \quad K_\perp' = K_\perp^2
\label{eq:RG2D}
\end{align}
Here, $Q=\tah(K)$ if $|\nu_c|>|\nu_s|$ and $Q=1/\tah(K)$ otherwise with $\nu_c$ and $\nu_s$ the eigenvalues of the 1D T-matrix, see  Eq.~(\ref{eq:1DTmatrix}). For $K_\perp < 1$ initially, $K_\perp^*=0$ is always approached implying that $K_\perp$ is an irrelevant perturbation. As a consequence, the fixed point describes a set of uncoupled 1D chains. Indeed, $\lambda(t)$ displays kinks, confirm Fig.~\ref{fig:2}. In this context let us therefore introduce an effective dimension $d^*$ which takes a value $d^*=1$ for the 1D system and also for the strongly anisotropic 2D one.

Decreasing the anisotropy makes the RG transformation in Eq.~(\ref{eq:RG2D}) less and less controlled. In particular, the isotropic point $J_\perp=J$ and its associated logarithmic singularity of Eq.~(\ref{eq:gScaling2D}) is not accessible in this way. Thus, within the current methodology, a rigorous identification of this DQPT with a fixed point is not possible. The particular scaling form of  $g(t)$ in Eq.~(\ref{eq:gScaling2D}), however, suggests that the DQPT in the isotropic limit is controlled by the unstable fixed point of the 2D Ising model. A further argument supporting this hypothesis is given below when discussing the power-law scaling of the spin correlations.


\begin{figure}
\centering
\includegraphics[width=\linewidth]{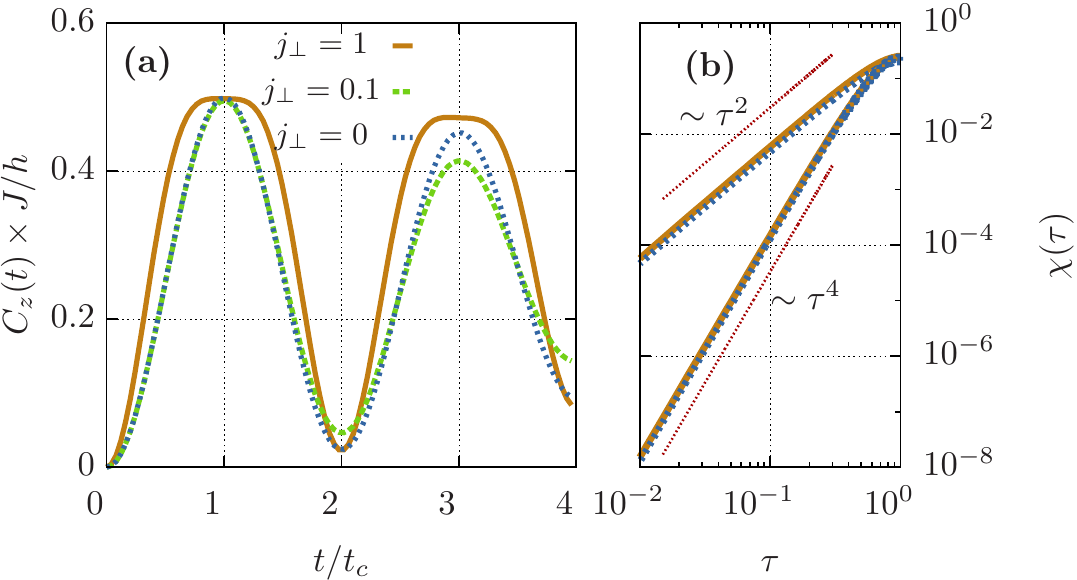}
\caption{(color online) (a) Dynamics of $C_z(t)$ for $h/J =0.1$ on a $N=5\times 5$ square lattice for different anisotropies $j_\perp$. (b) Power-law scaling of $\chi(\tau)$ in the vicinity of the DQPT. Brown curves are on the NN square lattice for $j_\perp=1/8$ with $d^\ast=1$ (upper curve) and the isotropic limit  $j_\perp=1$ with $d^\ast=2$ (lower curve). The dashed blue curves for the long-range Ising models relevant for trapped ions in 1D (upper curve) and in 2D (lower curve) are also included. As a reference, quadratic and quartic power-laws are also shown demonstrating the dynamical scaling of Eq.~(\ref{eq:dCz}).}
\label{fig:3}
\end{figure}
\emph{Spin correlations:--} Having unraveled scaling and universality of the Loschmidt amplitude, the major challenge now is to connect to local observables. Although for particular quenches out of symmetry-broken phases such a connection has been established~\cite{Heyl2014}, a general understanding, however, is still lacking. As the present DQPTs are associated with the equilibrium critical points of Ising models, it is a question of fundamental importance of how the divergent equilibrium correlation length becomes manifest in the nonequilibrium dynamics. Due to Lieb-Robinson bounds it is, of course, not possible to build up  diverging correlations within a finite time interval. Nevertheless, it will now be demonstrated that the underlying DQPTs are responsible for the buildup of NN spin correlations within the rows: 
\be
	C_z(t) = \frac{1}{L} \sum_{l m}  \langle \sigma_{l,m}^z(t) \sigma_{l,m+1}^z(t)\rangle.
\label{eq:Cz}
\ee
In Fig.~\ref{fig:3} the dynamics of $C_z(t)$ is shown.  Notice that the case $h=0$ constitutes a singular limit because there $C_z(t)$ is a constant of motion such that $C_z(t) \propto h/J$ for $h/J\ll 1$ and $C_z(t) J/h$ becomes a universal function independent of $h$ in the limit $h \to 0$. As one can see, the spin correlations develop a maximum in the vicinity of the DQPT at $t_c=\pi/(4J)$ with a slope, however, that differs for the isotropic 2D case compared to those with effective dimension $d^*=1$. This is associated with a different power-law scaling that can be quantified via:
\be
	\chi(\tau) = \frac{J}{h} \left[C_z(t_c)-C_z(t_c+\tau) \right]  \stackrel{h\to 0}{\propto} \tau^{2d^*},
\label{eq:dCz}
\ee
see Fig.~\ref{fig:3}. This scaling only depends on the effective dimension $d^*$ of the DQPT and therefore only on the DPQT's universality class if we assign $d^*=2$ for the DQPT satisfying the scaling of Eq.~(\ref{eq:gScaling2D}) for $g(t)$ equivalent to the critical point of the 2D Ising model. The numerical data in Fig.~\ref{fig:3} has been obtained from exact diagonalization (ED) using a Lanczos algorithm with full reorthogonalization~\cite{Cullum2002}.

\emph{Trapped ions:--} This dynamical scaling can be observed experimentally in systems of trapped ions within the current technology as will be outlined in the following. Fully polarized states as required in Eq.~(\ref{eq:initialState}) can be initialized with a high fidelity~\cite{Lanyon2011}. Coherent time evolution of transverse-field Ising Hamiltonians has been demonstrated both for 1D~\cite{Lanyon2011,Jurcevic2014,Richerme2014} and 2D\cite{Britton2012}. The Ising couplings, however, are not of NN type but rather long-ranged, $J_{lm}=J/|r_l-r_m|^\alpha$ with $0 \leq \alpha \leq 3$~\cite{Islam2013} and $r_l$ the location of the ion in real space.  In Fig.~\ref{fig:3}, numerical ED data of $\chi(\tau)$ for the long-range potentials is included for the dipolar case $\alpha=3$ with lattice spacing $a=1$. Here, open boundary conditions have been used and $\chi(\tau)$ includes only those NN correlations which do not contain spins at the boundary to minimize boundary effects. As the simulations indicate, the spin correlations also obey the dynamical scaling of Eq.~(\ref{eq:dCz}) making it accessible within current trapped ion technology.


\emph{Conclusions:--} It has been shown that DQPTs in 1D and 2D Ising models are controlled by unstable fixed points of complex RG transformations opening the possibility to apply the concepts of scaling and universality to the out-of-equilibrium regime. Importantly, this leads to a dynamical scaling of the spin correlations which only depends on the universality class of the underlying DQPT and which is accessible experimentally in systems of trapped ions.

\begin{acknowledgments}
Discussions with Philipp Hauke and Jan Budich are greatefully acknowledged. This work has been supported by the Deutsche Akademie der Naturforscher Leopoldina under grant numbers LPDS 2013-07 and LPDR 2015-01. 
The ED code uses the armadillo linear algebra libraries~\cite{Sanderson2010}.
\end{acknowledgments}


\bibliographystyle{apsrev}
\bibliography{literature}



\clearpage
\onecolumngrid
\setcounter{equation}{0}
\setcounter{figure}{0}

\begin{center}
{\bf \Large 
Supplemental Material to\vspace*{0.3cm}\\ 
\emph{Dynamical quantum phase transitions as critical points of complex renormalization group transformations}
}
\end{center}

\vspace*{0.3cm}
{\center{
\hspace*{0.1\columnwidth}\begin{minipage}[c]{0.8\columnwidth}
In this supplemental material, we provide methodological details for the derivation of the real-space RG equations. 
\end{minipage}
}
}

\section{A. Real-space RG in 1D: Decimation}
\label{app:sec:Decimation}

In the following, the well-known real-space decimation scheme for the one-dimensional Ising model is recapitulated, see Refs.~\cite{Nelson1975zz,Damgaard1993,Wei2014pt}. This will be the basis for the analysis below. Consider the (complex) partition function $Z(K,N)$ of the one-dimensional Ising model:
\be
	Z(K,N) = \mathrm{Tr} \left[ e^{\mathcal{H}(K,N)}\right],\quad \mathcal{H}(K,N) =K\sum_{l=1}^N \sigma_l^z \sigma_{l+1}^z
\ee
of length $N$ with $K$ a complex coupling in general, see main text. For simplicity, periodic boundary conditions, i.e., $\sigma_{N+1}^z=\sigma_1^z$, are used. Within decimation, it is the aim to eliminate every second spin yielding $Z(K,N) = Z(K',N/2)=\mathrm{Tr}\exp[\mathcal{H}(K',N/2)]$ with a new Hamiltonian $H(K',N/2)$ for the remaining $N/2$ spins that now interact via the renormalized coupling $K'$. As the Pauli matrices $\sigma_l^z$ mutually commute, the partition function can be factorized such that integrating out all the odd spins gives:
\be
	Z(K,N) = \mathrm{Tr}_e \mathrm{Tr}_o \prod_{l=1}^N e^{K\sigma_{l}^z \sigma_{l+1}^z} = \mathrm{Tr}_e \prod_{l=1}^{N/2} \sum_{\sigma_{2l+1}^z=\pm 1} e^{K\sigma_{2l}^z\sigma_{2l+1}} e^{K\sigma_{2l+1}^z\sigma_{2l+2}}
\ee
with $\mathrm{Tr}_{e/o}$ denoting the trace over the even (e) or odd (o) sites, respectively. Using $\exp[K\sigma \sigma']=\cosh(K) + \sinh(K) \sigma\sigma'$ for $\sigma$ and $\sigma'$ Pauli matrices, one obtains:
\be
	\sum_{\sigma_{2l+1}^z=\pm 1} e^{K\sigma_{2l}^z\sigma_{2l+1}^z} e^{K\sigma_{2l+1}^z\sigma_{2l+2}^z} = \cosh^2(K) \left[ 1+ \tanh^2(K) \sigma_{2l}^z\sigma_{2l+2}^z \right],
\ee
This gives the desired RG equation for the couplings:
\be
	\tanh(K') = \tanh^2(K), \quad e^{2K'} = \cosh(2K).
\ee
Having recapitulated the decimation RG scheme for the classical 1D Ising chain, it is the aim of the following to study the influence of weak transverse fields.

\section{B. Weak transverse fields}
\label{app:sec:Fields}

In this section the perturbative calculations for the Ising chain in presence of transverse fields are shown. First, the mapping onto the classical Ising with next-to-nearest neighbor interactions is presented using time-dependent perturbation theory. Afterwards, the RG equations of the main text are derived on the basis of standard cumulant expansion.

\subsection{Mapping onto classical Ising chain}

The Ising chain with a transverse field $h$ reads:
\be
	H = H_0  + V, \quad H_0 = -J\sum_l \sigma_l^z \sigma_{l+1}^z,\,\, V= -h \sum_l \sigma_l^x.
\ee
In order to determine the Loschmidt amplitude $\mathcal{G}(t) = \langle \psi_0 | \exp[-iHt] | \psi_0 \rangle$, see Eq.~(1) of the main text, for a weak transverse field, let us turn to an interaction picture with respect to $H_0$:
\be
	e^{-iHt} = e^{-iH_0t} W(t),\quad W(t)= \mathcal{T} e^{-i\int_0^t \mathrm{d}t' \,\, V(t')}, \quad V(t) = e^{iH_0t} V e^{-iH_0 t},
\ee
with $\mathcal{T}$ the usual time-ordering prescription. For the Loschmdit amplitude this implies:
\be
	\mathcal{G}(t) =  \langle \psi_0 |e^{-iH_0t} W(t) | \psi_0 \rangle = \sum_{\mathbf{s}} e^{-iE_\mathbf{s} t} \langle \psi_0 | \mathbf{s}\rangle \langle \mathbf{s}|W(t) | \psi_0 \rangle, \quad \langle \mathbf{s} |W(t)|\psi_0\rangle = \langle \mathbf{s}|  \mathcal{T} e^{-i\int_0^t \mathrm{d}tV(t')} |\psi_0\rangle.
\ee
by inserting an identity via $1=\sum_\mathbf{s} |\mathbf{s}\rangle \langle \mathbf{s}|$ with $|\mathbf{s}\rangle = |s_1,\dots,s_L\rangle$ denoting the complete basis of spin configurations and $s_l=\pm 1$ the orientation of the local spin at site $l$. Using standard cumulant expansion methods for time-ordered exponentials~\cite{Kubo1962xh} one obtains to lowest order in the transverse-field strength $h$:
\be
	\langle \mathbf{s}|  \mathcal{T} e^{-i\int_0^t \mathrm{d}t} |\psi_0\rangle \approx \langle\mathbf{s} |\psi_0\rangle \exp\left[ -i\int_0^t \mathrm{d}t' \frac{\langle \mathbf{s}|V(t')|\psi_0\rangle}{\langle\mathbf{s}|\psi_0\rangle} \right]
\ee
For the choice $|\psi_0\rangle$ as the ground state of $V$ of the main text this gives
\be
	\langle \mathbf{s} |  \mathcal{T} e^{-i\int_0^t \mathrm{d}t} |\psi_0\rangle = 2^{-L/2} \exp[h(1-\cos(4Jt))/(4J)\sum_l s_l s_{l+1} -i \frac{h}{2} (t-\sin(4Jt)/(4J))\sum_{l} s_l s_{l+2}]
\ee
This result can be rewritten in terms of a classical Ising model but now including also next-to-nearest neighbor (NNN) interactions which in the end gives for the Loschmidt amplitude the desired result:
\be
	\mathcal{G}(t) = \mathrm{Tr} e^{\overline{H}}, \quad \overline{H} = K \sum_l \sigma_l^z\sigma_{l+1}^z + G\sum_l \sigma_l^z \sigma_{l+2}^z,
\ee
with
\be
	K = iJt + \frac{h}{4J} \left[1-\cos(4Jt) \right],\quad G = -i\frac{h}{2}t + i\frac{h}{8J}\sin(4Jt).
\ee

\subsection{RG equations}

It is the aim of the following to apply the real-space decimation approach to the Ising chain with NNN interactions. Again, we will eliminate every second spin, which we take to be odd lattice sites without loss of generality:
\be
	e^{\overline{H}'} = \mathrm{Tr}_o e^{\overline{H}}, \quad \overline{H}=\overline{H}_0 + \overline{V},\quad \overline{H}_0 = K \sum_l \sigma_l^z\sigma_{l+1}^z, \,\, \overline{V} = G\sum_l \sigma_l^z \sigma_{l+2}^z.
\ee
Here, $\overline{H}'$ denotes the renormalized Hamiltonian after this RG step for the remaining even lattice sites and $\mathrm{Tr}_o$ is the trace over all the odd lattice sites that are supposed to be integrated out. As $[\overline{H}_0,\overline{V}]=0$ it holds that
\be
	e^{\overline{H}'} = \mathrm{Tr}_o \left[ e^{\overline{H}_0} \right] \frac{\mathrm{Tr}_o \left[e^{\overline{H}_0} e^{\overline{V}} \right]}{\mathrm{Tr}_o \left[e^{\overline{H}_0} \right]} = \mathrm{Tr}_o \left[ e^{\overline{H}_0} \right]  \langle \langle  e^{\overline{V}} \rangle \rangle,
\ee
with $\langle\langle \dots \rangle\rangle$ denoting a generalized average. As $\overline{V}$ is proportional to $G$, which is assumed to be small, one can perform a cumulant expansion~\cite{Niemeyer1973tr} which to lowest order is
\be
	\langle \langle  e^{\overline{V}} \rangle \rangle \approx \exp\left[ \langle \langle  \overline{V} \rangle \rangle \right] 
\ee
Evaluating $\langle \langle  \overline{V} \rangle \rangle$ yields the desired RG equations presented in the main text.

\section{C. Anisotropic 2D Ising model}
\label{app:sec:2D}

The anisotropic 2D Ising model considered in the main text reads:
\be
	H = -J\sum_{lm} \sigma_{l,m}^z \sigma_{l,m+1}^z - J_\perp \sum_{lm} \sigma_{l,m}^z \sigma_{l+1,m}^z
\ee
with $l$ ($m$) denoting the row (column) on the 2D square lattice. The coupling within the rows is of strength $J$ whereas along the columns it is $J_\perp$. In the following, a perturbative real-space RG will be derived taking $j_\perp=J_\perp/J$ as a small parameter. Each isolated row is an exactly solvable 1D Ising model. It is the aim the eliminate every second (odd) row. For the Loschmidt amplitude $\mathcal{G}(t)$ one can then introduce the decomposisiton
\be
	\mathcal{G}(t) = \mathrm{Tr} \left[ e^{\mathcal{H}} \right],\quad \mathcal{H} = \mathcal{H}_e + \mathcal{H}_o + \mathcal{H}_{eo}
\ee
with
\be
	\mathcal{H}_e = K\sum_{lm} \sigma_{2l, m}^z \sigma_{2l,m+1}^z,\quad \mathcal{H}_o = K\sum_{lm} \sigma_{2l+1,m}^z \sigma_{2l+1,m+1}^z,\quad \mathcal{H}_{eo} = K_\perp \sum_{lm} \sigma_{l,m}^z \sigma_{l+1,m}^z.
\ee
The elimination of the odd rows can be formally implemented in the following way:
\be
	e^{\mathcal{H}'} = e^{\mathcal{H}_e} \mathrm{Tr}_o \left[e^{\mathcal{H}_o} \right] \frac{\mathrm{Tr}_o\left[e^{\mathcal{H}_o} e^{\mathcal{H}_{eo}} \right] }{\mathrm{Tr}_o\left[e^{\mathcal{H}_o} \right] } = e^{\mathcal{H}_e} \mathrm{Tr}_o \left[e^{\mathcal{H}_o} \right] \langle\langle e^{\mathcal{H}_{eo}} \rangle\rangle.
\ee
Because $\mathcal{H}_{eo}$ is proportional to the weak coupling $K_\perp \ll 1$ it is again possible to perform a cumulant expansion:
\be
	\langle\langle e^{\mathcal{H}_{eo}} \rangle\rangle \approx \exp\left[ \frac{1}{2}\langle\langle \mathcal{H}_{eo} \mathcal{H}_{eo} \rangle \rangle \right]
\ee
which now has to be determined up to second order because the first order vanishes. Neglecting the perturbatively generated longer-ranged interactions one obtains the desired RG equations 
\be
	K' = K + 2 Q K_\perp^2,\quad K_\perp' = K_\perp^2,
\ee
with $Q$ the nearest-neighbor correlation function of the 1D Ising model which is $Q=\tanh(K)$ if $|\nu_c|>|\nu_s|$ and $Q=1/\tanh(K)$ otherwise~\cite{Sachdev2011}.

\end{document}